\documentclass[preprint2]{aastex}

 \usepackage{lineno}
 \linenumbers*[1]

\shorttitle{CME $\beta$ transition}
\shortauthors{Savani et al.}

\graphicspath{{figure1/}}

\begin{document}
\title{A magnetic boundary layer creating a quasi-cylindrical substructure within a propagating flux rope leading to a plasma beta transition}

\author{N.P. Savani \altaffilmark{1}}
\affil{George Mason University, Faifax, VA, USA}
\author{A. Vourlidas}
\affil{Space Science Division, Naval Research Laboratory, Washington, DC 20375-5352, USA}
\author{D. Shiota}
\affil{Computational Astrophysics Laboratory, Advanced Science Institute, RIKEN, 2-1 Hirosawa, Wako, Saitama 351-0198, Japan}
\author{M. G. Linton}
\affil{Space Science Division, Naval Research Laboratory, Washington, DC 20375-5352, USA}
\author{K. Kusano \altaffilmark{2}}
\affil{Solar-Terrestrial Environment Laboratory, Nagoya University, Nagoya 464-8601, Japan}
\author{N. Lugaz}
\affil{Space Science Center and Department of Physics, University of New Hampshire, Durham, NH 03824, USA}
\and
\author{A.P. Rouillard \altaffilmark{3}}
\affil{Institut de Recherche en Astrophysique et Plantologie, Universit de Toulouse (UPS), France}

\altaffiltext{1}{NASA Goddard Space Flight Center, Greenbelt, MD 20771, USA}
\altaffiltext{2}{Japan Agency for Marine-Earth Science and Technology, Yokohama, Kanagawa 236-0001, Japan}
\altaffiltext{3}{UMR 5277, Centre National de la Recherche Scientifique, Toulouse, France}


\begin{abstract}
We present a 2.5D MHD simulation of a magnetic flux rope (FR) propagating in the heliosphere and investigate the cause of the observed sharp plasma $\beta$ transition. Specifically, we consider a strong internal magnetic field and an explosive fast start, such that the plasma $\beta$ is significantly lower in the FR than the sheath region that is formed ahead. This leads to an unusual FR morphology in the first stage of propagation, while the more traditional view (e.g. from space weather simulations like Enlil) of a `pancake' shaped FR is observed as it approaches 1AU. We investigate how an equipartition line, defined by a magnetic Weber number, surrounding a core region of a propagating FR can demarcate a boundary layer where there is a sharp transition in the plasma $\beta$. The substructure affects the distribution of toroidal flux, with the majority of the flux remaining in a small core region which maintains a quasi-cylindrical structure.  Quantitatively, we investigate a locus of points where the kinetic energy density of the relative inflow field is equal to the energy density of the transverse magnetic field (i.e. effective tension force). The simulation provides compelling evidence that at all heliocentric distances the distribution of toroidal magnetic flux away from the FR axis is not linear; with $80\%$ of the toroidal flux occurring within $40\%$ of the distance from the FR axis. Thus our simulation displays evidence that the competing ideas of a pancaking structure observed remotely can coexist with a quasi-cylindrical magnetic structure seen in situ.

\end{abstract}


\section{Introduction}

Coronal mass ejections (CMEs) are large scale transient eruptions from the Sun that travel into the heliosphere. The impact of these transients onto Earth are the dominant source of the most severe space weather incidents \citep{tsurutani1997}. The velocity and magnetic field vectors within a CME are often considered the primary properties of interest for estimating their space weather effects. CMEs also play a significant role in removing magnetic helicity which is generated as the Sun evolves \citep[e.g.][]{demoulin2002}. It has also been shown to provide a significant contribution in modulating the open solar flux over a solar cycle \citep[e.g.][]{owens2008}. This open solar flux has been correlated to the long term variations of the heliospheric magnetic field \citep[e.g.][]{barnard2011,lockwood2011}. Therefore understanding the distribution and amount of toroidal flux contained within CMEs are important for estimating the severity of a space weather incident as well as for long term solar influence on Earth's climate.

The morphology of a CME containing a magnetic flux rope (FR) during its propagation towards Earth is often described as `pancaking' \citep{riley2004}. Historically, this is supported by remote observations from coronagraphs which display CMEs propagating radially away from the Sun, and more recently with heliospheric imagers on board STEREO \citep{howard2008}. Also, these cameras often display a flux rope structure as a black void within the images such that the observed white light marks the outer boundary, such as sheath regions \citep[e.g.][]{howardt2012b,deforest2013}. Synthetic white light images from simulations with a magnetic flux rope have also been shown to produce a dark cavity \citep{lugaz2005}. These observations are also further supported by simulations that are often used in space weather forecasting (e.g. Enlil model). The Enlil model describes a CME as a high density packet without a coherent magnetic field structure \citep{odstrcil1999}. Both the STEREO observations and the Enlil model display a CME uniformly stretching perpendicular to the radial flow, while contradictory evidence is often presented with in situ measurements. Flux rope models that satisfy the Lundquist solution \citep[e.g.][]{lepping1990,savani2013} are often successfully shown to reproduce similar results to the magnetic vectors from in situ measurements within CMEs. The morphological structure of this Lundquist-solution flux rope model is cylindrical with a circular cross section (i.e. not stretched into a `pancaking' structure).  In this paper we present evidence, through the use of simulations, that a CME can display both the `pancaking' properties seen remotely and circular cross section measured in situ.


Flux ropes are often characterised by magnetic fields that are stronger than the ambient solar wind and are topologically defined to have a coherent rotation of their magnetic field lines around a central axis. These flux rope CMEs have often been observed to possess a relatively low proton temperature when detected in situ, and were labelled magnetic clouds \citep[M.C.,][]{burlaga1981}. The fraction of CMEs that contain a FR structure is under debate, with statistical arguments suggesting that there is a solar cycle dependence as well there being a variation between $30\%$ \citep{richardson2010} and $>70\%$ \citep{marubashi2000}. Recently, further evidence suggests that all CMEs may in fact be associated with a magnetic flux rope topology by hypothesising certain events without these signatures are due to the spacecraft trajectory or a misidentification \citep{vourlidas2012}.

A successfully simple technique to model the magnetic field structure of flux ropes from solar wind in situ measurements has been a force-free model. Some of the earliest attempts assumed a cylindrically symmetric condition \citep{lepping1990,marubashi1997} and due to their success, continue to be used to this day \citep[e.g.][]{dasso2005, savani2013}. They used a constant alpha value in the Bessel function solutions to $ \nabla \times \mathbf{B}= \alpha \mathbf{B}$ \citep{lundquist1950}. Further developments to these models have included ad hoc expansion to the field vectors \citep{marubashi1997}, incorporating an elliptical cross section \citep{hidalgo2002,owens2006} and curved flux rope axis \citep{marubashi2000, owens2012}.

An alternative technique to the modelling has been developed by \citet{hu2002}. By using magnetic field and plasma measurements and integrating the nonlinear, planar Grad-Shafranov (GS) equation, the technique is able to estimate some global CME properties similar to that of flux rope models mentioned earlier. Unlike the models, the GS technique does not enforce any fixed global morphology for the 2D cross-section of the flux rope. This leads to estimates of magnetic flux and flux rope size that are not constrained to a predetermined model morphology. Both of these parameters are pertinent in this study.

The popularity and success of the cylindrical flux rope has often led research to progress by implementing an idealised morphology as a first approximation, e.g. in the case of a 3D reconstruction technique for coronagraphs \citep{thernisien2009, thernisien2011}, or for the growth rate of the cross-sectional width further in the heliosphere \citep{savani2009}. Extensions to these studies have been carried out to better understand the distortions towards a more elliptical shape as the CME propagates into the heliosphere \citep{nieves2012, schreiner2012}. However the variations in the ambient solar wind can further develop the morphology into a more complex structure \citep[e.g.][]{odstrcil1999, owens2006b, liu2006, savani2010}.

CMEs are often considered to propagate radially away from the Sun. This behaviour forces an expansion onto the object that is perpendicular to its radial propagation. This effective velocity would usually be different from the internal radial expansion which is commonly attributed to a pressure force. A kinematic study by \citet{riley2004} inferred that a CME would drastically deviate away from the force-free cylindrical configuration and labelled the expansion as `pancaking'. However a more detailed investigation into the aspect ratio of the CME cross section by \citet{savani2011a} revealed that the global morphology of an initially circular structure would approach a steady state. If the radial expansion is considered to equal a fixed fraction of the bulk flow then the aspect ratio would tend towards a constant value of $\sim5$, depending on the original size of the CME.

However, the success of the circular FR fits may be due to the CME structure being much more circular on a local scale. This suggests that at certain locations, the magnetic tension or external forces due to structured solar wind conditions could restore the magnetic vectors towards the minimum energy configuration of the Bessel functions \citep{suess1988}. A study by \citet{savani2011b} supports this scenario as the authors investigated the shock stand-off distance of CMEs to infer the aspect ratio. The results from 45 CMEs gave an average aspect ratio of $2.8\pm0.5$. Also, a visual inspection of the several studies with reconstructed CME morphology using the GS technique have shown ratios consistently $<4$ \citep[e.g.][]{liu2008, mostl2009a, wood2012b}.

Estimates of the total flux content within CMEs have important consequences for estimating the long term trends of the total open solar flux \citep{lockwood2004, owens2006c}. The study by \cite{owens2006c} assumed the heliosphere contained a constant background of open flux with a time-varying component due to interplanetary CMEs. The contribution of CME flux was estimated from 132 magnetic clouds with the analysis assuming a constant-$\alpha$ force free FR with a circular cross-section \citep{lynch2005}. This paper tests the validity of assuming a circular cross-section for estimating the flux content as it is often the assumption used \cite[e.g.][]{blanco2013}.


\section{MHD simulations}
We have implemented an axisymmetric MHD simulation of a flux rope CME propagating out to 300 Solar radii (Rs). The ambient solar wind is initialised with an inner boundary at 4Rs and propagated at a uniform supersonic speed of 300km/s. The magnetic flux rope is super-imposed at 10 Rs as a semi-analytically calculated toriodal solution to the azimuthally symmetric Grad-Shafranov equation (where $\phi$ is defined as the toroidal direction and parallel to the FR axis). Initially the FR is given a radial width of 8Rs with a speed of 700 km/s. In Figure 1, we summarize the evolution of the flux rope at three intervals. The green line indicates the boundary of the FR and the black contours display magnetic field lines which follow the flux function, $\Psi$, defined by
\begin{equation}
\mathbf{B_{\perp}}=\mathbf{\nabla} \times 
\left( \begin{array}{c} \frac{\Psi}{r \sin \theta}\  \mathbf{\hat{\phi}}  
\end{array} \right) ,
\end{equation}

where $\mathbf{B_{\perp}}$ is the in-the-plane magnetic field. Detailed mathematical formulation of the simulation is presented elsewhere \citep{shiota2008,savani2012b} and we restrict our discussion to a brief description of the initial conditions and the evolution of the FR. 

The simulated FR is placed into the heliospheric region with an instantaneous velocity equal to 700 km/s. While this does not capture the initiation and low coronal phases of the CME evolution, it is a reasonable approximation for fast CMEs that often originate in active regions. The flux rope was initiated with a toroidal (axial, or out-of-the-plane) flux of $3 \times 10^{21}$Mx which is of a typical order of magnitude observed. Figure 1 displays the plasma $\beta$ on the colour scale to highlight the regions that are more magnetically dominated than others. The plasma $\beta$ is defined as the ratio of the plasma pressure $(P = n k_B T)$ to the magnetic pressure $(P_{mag} = \mathbf{B}^2/2\mu_0)$.  The temperature and density of the background solar wind are not replicated accurately relative to observations, as such, the quantitative values in the plasma $\beta$ is therefore not realistic (i.e. at 1 AU the simulation density and temperature are larger than observed). However, the focus of this investigation is the relative changes in the plasma $\beta$ from one region to the next, which is expected to be well reproduced by the simulation.

\section{Results}

\subsection{Plasma $\beta$ Transition}
\begin{figure}
\includegraphics[scale=0.45]{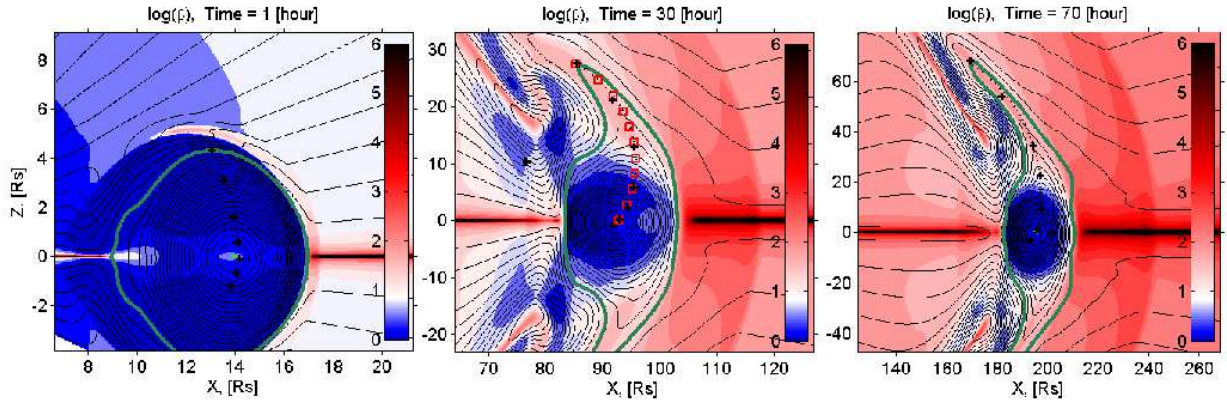}
\caption{The plasma $\beta$ (colour scale) and magnetic flux function(black contours) are displayed for three times during an MHD calculation. The edge of the closed field lines for the magnetic flux rope are displayed as a thick line(green). The black crosses denote manually selected positions through the flux rope, and the dashed line is an arc of a circle that is optimally chosen from the crosses. The middle panel identifies the uniformly distributed positions along the arc that is later used in our analysis (red squares). The figure shows that as the flux rope propagates the outer layers increase in plasma $\beta$.}
\label{beta}
\end{figure}

The entire FR begins with low plasma $\beta$ and its fast release quickly produces a leading shock front ahead. Figure \ref{beta} shows that within the sheath region ahead of the FR leading edge, the plasma $\beta$ increases significantly above both the ambient solar wind and FR values. From the beginning of the FR propagation, the plasma dominated sheath region can be seen to drag the open field lines locally around FR. After some time, these regions also interact with the outer edges of the FR such that the plasma $\beta$ in these areas also increases. These outer `layers' of the FR are then also dragged away from the ecliptic plane with the ambient solar wind (see also Figure 3 in \citet{savani2012b}).

By the time the FR approaches 215Rs ($\sim1AU$), the outer boundary of the flux rope (Figure \ref{beta}, green curve) more closely resembles an evenly pancaked structure (as shown by \citet{riley2004}) than earlier ($\sim80Rs$) in its propagation. The FR edge is defined by a boundary between closed and open field lines. Upstream of the FR is a sheath region containing `shocked' solar wind material which is downstream of the shock wave. In the early evolution of the flux rope, the outer boundary appears closer to a circular substructure with extended `wing' segments. The formation of a wake is quickly established downstream of the `wings'. In our case, similar to the case shown by \citet[][hereinafter referred to as ``EM98"]{emonet1998}, the outer layers of the flux rope with weaker magnetic field are ``unable to resist the incipient external flow, so that they are bodily convected to the rear".

The plasma $\beta$ within the outer boundary is predominately below ten in the initial conditions as well as in the early evolution, whereas the sheath region is significantly higher. However, as the flux rope propagates the higher plasma $\beta$ of the sheath region smoothly changes to a lower value in the CME core.  

\subsection{Estimating the Position of the Substructure}

\begin{figure}
\includegraphics[scale=0.80]{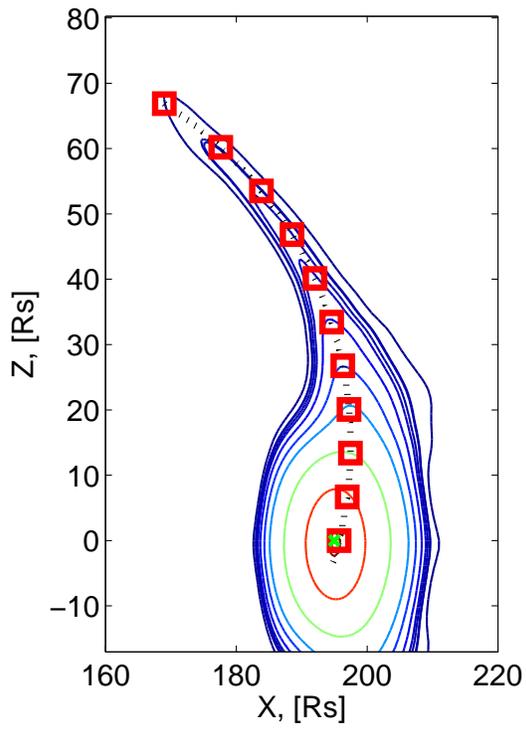}
\caption{Contours of constant magnetic flux function are drawn for the 11 equidistant positions (red squares) along the optimally chosen arc (black dashed line, see Figure \ref{beta}) from within the flux rope.}
\label{fluxcont}
\end{figure}

We see that the concentrated magnetic flux seen in the earlier evolution of the FR still persists out to 1 AU. In order to investigate the distribution of flux, we manually selected an arc between the outer most edge of the FR (green curve) to the center as defined by the maximum flux function (green cross). We choose to define the points to approximately occur at the positions where the magnetic field line is a maximum distance away from the center. A circular arc was then created to best fit the selected data points (black dashed line in Figure \ref{beta} and \ref{fluxcont}). As this paper is interested in understanding how the flux is `stretched' in the out-of-ecliptic plane (Z direction), eleven positions along the arc were chosen to be uniformly distributed along the Z direction (Figure \ref{fluxcont}, red squares). The toroidal flux contained within the contours of these constant flux function values were then calculated (curves in Figure \ref{fluxcont}). The distance was then calculated as the position on the contour furthest away from the FR center. This method enables the flux content to be sampled at an approximately uniform rate with regards to the distance away from flux rope center. Other methods were investigated, but they often concentrated the points within the inner core of the flux rope with the outer edge of the flux rope being under-sampled. Also, this procedure meant that the exact positions from the initial manual selections does not matter as the maximum distance of the field line and the out-of-plane flux contained within the closed loop are estimated independently. 

\begin{figure}
\includegraphics[scale=0.70]{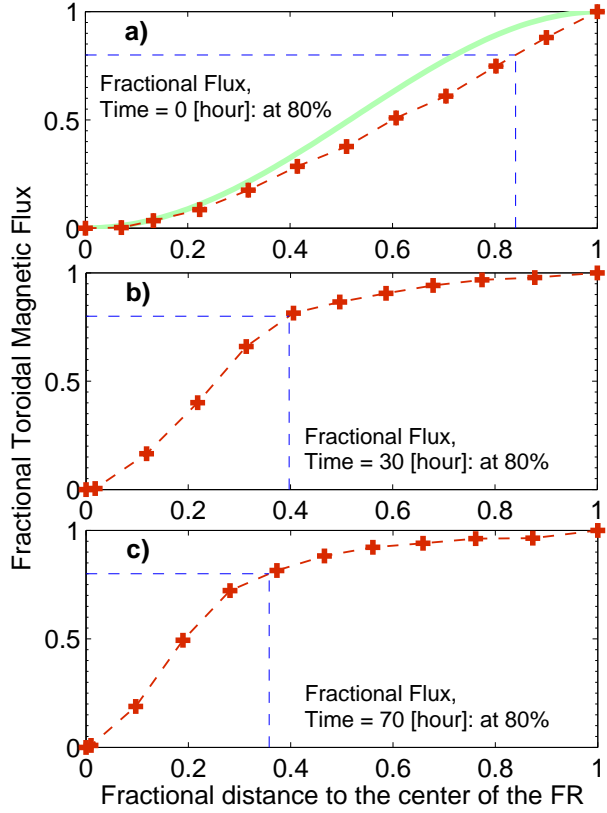}
\caption{The distribution of toroidal magnetic flux against the distance away from the center of the flux rope. Both the magnetic flux and size of the flux rope are normalised to the values of the outer edge. Top panel shows the distribution in the initial condition as well as the idealised cylindrical Lundquist solution (thick green curve). The red crosses are the magnetic flux values from positions chosen in Figure \ref{fluxcont}.  This figure shows that the majority of the toroidal magnetic flux is concentrated in the inner core of the whole flux rope. }
\label{fracFlux}
\end{figure}

Figure \ref{fracFlux} displays the distribution of toroidal flux as a function of the distance to the FR center. In order to better compare the results from different time frames, the toroidal flux values have been normalised by the maximum value at the outer edge ($\Phi(r)/\Phi_{0}$); and the distance away from the FR center was normalised by the maximum diameter of the FR ($r/R_0$). The top panel of Figure \ref{fracFlux} shows the initial distribution (t=0 hours) of the FR which is a solution of toriodal Grad-Shafranov equation. For comparison, the flux distribution from the cylindrical Bessel function is also displayed as a thick green curve. The flux distribution is clearly not uniform at $\sim100$ Rs, but interestingly the distribution remains non-uniform even when the morphology may appear otherwise. The results show that beyond 100Rs the majority of the flux ($80\%$) remains within an inner core that is about$~40\%$ of the overall size.

\begin{figure}
\includegraphics[scale=0.70]{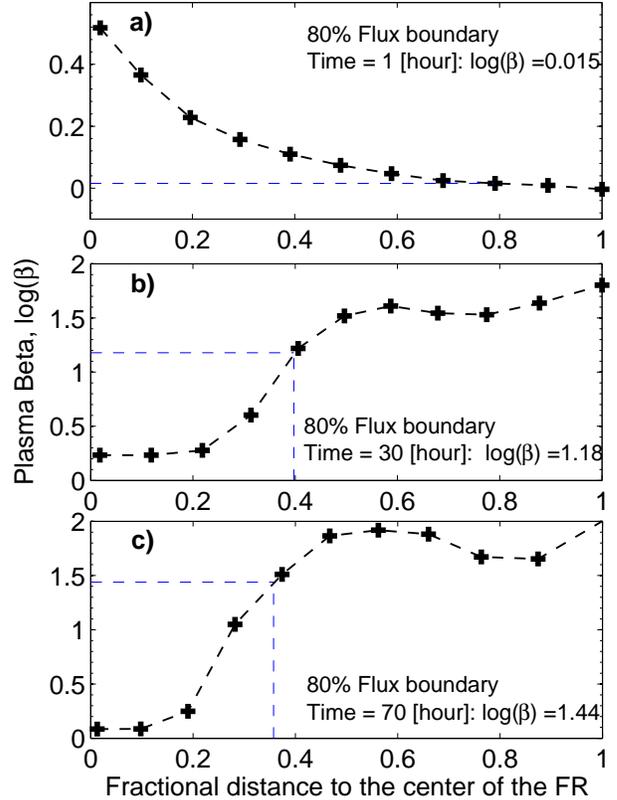}
\caption{The distribution of plasma $\beta$ against the distance away from the center of the flux rope. The flux rope is normalised to the values of the outer edge. The three panels represent the same time intervals displayed in Figure \ref{beta}. The black crosses are the $\beta$ values for the same positions shown in Figure \ref{fluxcont} and \ref{fracFlux}.This figure shows that the flux rope has a sharp transition in plasma $\beta$ which coincides with approximately 80 $\%$ of the toroidal magnetic flux.}
\label{betaProf}
\end{figure}

Figure \ref{betaProf} displays the plasma $\beta$ values for the eleven positions within each of the frames shown in figure \ref{beta}. A transition region between the more magnetically and plasma dominated regimes (low to high plasma $\beta$) is found at $\sim80\%$ of the total toroidal flux content. This boundary is shown as dashed blue lines in figure \ref{fracFlux} and \ref{betaProf}. The formation of these stretched high $\beta$ regions leads to a morphology of `wings' and the disconnected wake (vortical eddy's)to the rear.

\subsection{Magnetic Weber Number}
To understand the causality of the quasi-cylindrical magnetic core we investigate the source of the resistance to deformation which leads to the observed transition from low to high plasma $\beta$. The two competing forces acting on the periphery of the flux rope are: 1. the act of deformation which can be considered as the kinetic energy density of the relative flow onto the flux rope; 2. the act of restoration which is represented by the energy density of the poloidal magnetic field (i.e. the tension force of a twisted field).

\begin{figure}
\includegraphics[scale=0.64]{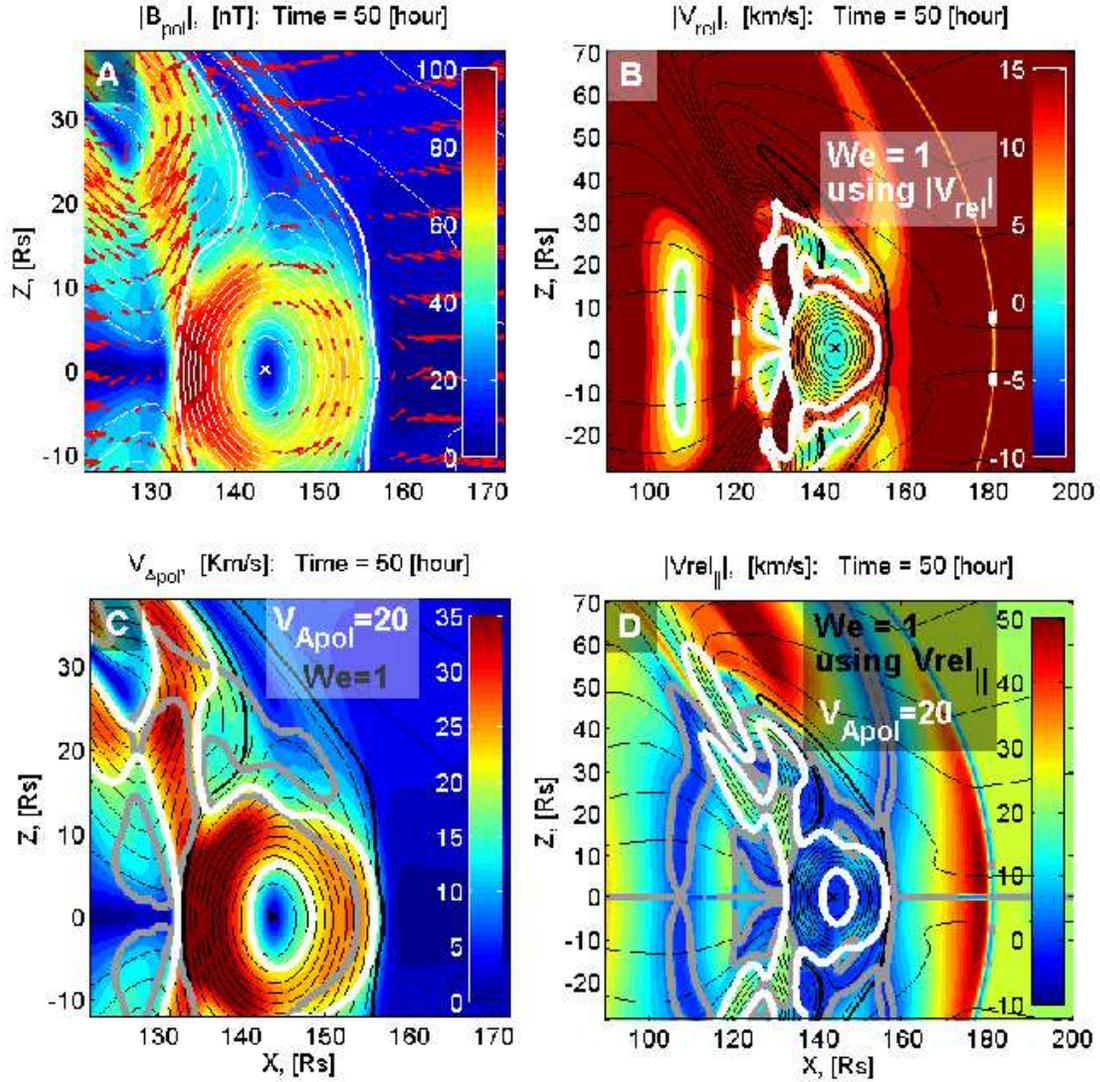}
\caption{Various properties of the FR is displayed at a heliocentric distance of $\sim145$ solar radii. (a) Displays the poloidal component of the magnetic field $|B_{pol}|$ (colour scale) and magnetic flux function (white contours) with the velocity vectors of the solar wind ($\mathbf{v}_{rel_{\parallel}}$). The $|\mathbf{v}_{rel_{\parallel}}|$ is the parallel component to the magnetic field and measured relative to the FR center. (b) Displays the magnitude of the velocity relative to the FR center (colour scale) with the Weber equipartition line, We, which was determined using $\mathbf{v}_{rel}$ similar to \citet{emonet1998} (white curve) (c) Displays the poloidal component of the Alfv\'{e}n speed (colour scale) with the We line (grey curve) and the contour of  $\mathbf{v}_{A_{pol}}= 20km/s$ (white curve). (d) Displays the $|\mathbf{v}_{rel_{\parallel}}|$ (colour scale) with the $We_{\parallel}$ equipartition line which was determined using $|\mathbf{v}_{rel_{\parallel}}|$ (grey curve) and the contour of  $\mathbf{v}_{A_{pol}}= 20km/s$ (white curve).}
\label{2by2}
\end{figure}

The kinetic energy density is defined by, $e_{kin} \equiv \rho v_{rel}^{2}/2$. Where $ \mathbf{v}_{rel}$ is the relative velocity of the solar wind onto the flux rope. The poloidal-field energy density is defined by $e_{mag} \equiv B_{pol}^{2}/8\pi$. Where $B_{pol}$ is the (in-the-plane) poloidal component of the magnetic field. Figure 5a displays a circular ring of high intensity $B_{pol}$ that is part of the core FR region. From the ring to the center of the FR, the $B_{pol}$ field strength decreases rapidly as the central region of the FR becomes more dominated by an axial field direction (toroidal direction). The contours of the flux function are displayed as white curves on Figure \ref{2by2}a, with a thicker curve marking the outer boundary of the FR as determined by a outermost closed field line. Other regions (out of the equatorial plane and to the rear of the FR) also have a high $B_{pol}$ field strength. These regions surround the wake that forms behind the FR `wings'.

We define the equipartition line by the same method as EM98; i.e. when the two energy densities are equal. This definition is also equivalent to identifying the locus of all points where the magnetic Weber number equals one. The magnetic Weber number is defined as:

\begin{equation}
We \equiv \frac{\mathbf{v}_{rel}^{2}\rho }{B_{pol}^{2}/ \left(4\pi \right) } ,
\end{equation}

The Weber number can therefore also be expressed in relation to the poloidal component of the Alfv\'{e}n velocity, $(\mathbf{v}_{A_{pol}})$:
\begin{equation}
We = \frac{\mathbf{v}_{rel}^{2}}{\mathbf{v}_{A_{pol}}^{2} } .
\end{equation}

In the hydrodynamical literature, the Weber number is a dimensionless unit that measures the relative importance of a fluid's inertia compared to its surface tension. In our magnetohydrodynamic case, the role of the surface tension is replaced by the magnetic tension of the poloidal field component.

EM98 studied the physics of a rising twisted magnetic flux tube in two dimensions, under conditions akin to the solar convection zone. In this situation, the flux tube rose slowly without the formation of a shock front and through a non-magnetic background medium. Under this scenario, it was appropriate to define the relative velocity as $ \mathbf{v}_{rel}\equiv \mathbf{v} - \mathbf{v}_{apex}$. Where $\mathbf{v}_{apex}$ is the velocity at a single location - the apex of the flux tube. By choosing the relative velocity with respect to the location of the flux rope center we display the Weber equipartition line (white curve) in figure \ref{2by2}b. The colour scale displayed in Figure \ref{2by2}b is the magnitude of the relative velocity. For the front and rear boundary of the flux rope, near to the equatorial plane, the Weber boundary accurately follows the drastic change in the plasma $\beta$. The Weber boundary reflects a quasi-circular morphology surrounding the core region of the FR. However, at the rear of the FR the Weber boundary bridges to a another region out of the equatorial plane. For regions significantly outside the equatorial plane, the Weber boundary as defined by EM98, inappropriately extends further into the `wing' regions of the flux rope. This region also extends significantly into the magnetically disconnected wake.

To better understand the complex nature of the Weber line, the contour of We=1 is over-plotted onto an intensity map of the $|\mathbf{v}_{A_{pol}}|$ (see figure \ref{2by2}c). Figure \ref{2by2}c displays the Weber contour (grey curve) with an arbitrarily chosen contour of $|\mathbf{v}_{A_{pol}}| =20 km/s$ (white curve). With regards to the core of the FR, both boundaries display very similar quasi-circular morphology that remain within the closed magnetic field lines of the FR. The extended regions of the Weber line that stretch into the wake also occurs for the $|\mathbf{v}_{A_{pol}}|$ line. The bridge that connects to the extended regions out of the equatorial plane appears to be caused by a high poloidal magnetic field strength. Whereas, the extended regions themselves seem to be caused by a reduction in the relative flow velocity ($\mathbf{v}_{rel}$).

A significant difference between the study performed by EM98 and the one presented here is the direction of the solar wind flow falling onto the leading edge of the flux rope. The presence of a leading shock front deflects the solar wind within the sheath region prior to arriving at the flux rope leading edge. The presence of a shock front clearly adds further complication to our scenario. Therefore to correct for the Weber equipartition line, we re-define the relative velocity to equal the component parallel to the in-plane magnetic field and relative to flux rope center. i.e we define it as

\begin{equation} \label{eqVe}
\mathbf{v}_{rel_{\parallel}} = \frac{\mathbf{B_{pol}}\cdot(\mathbf{v}-\mathbf{v}_{center})}{|\mathbf{B_{pol}}|^2}  \cdot   \mathbf{B_{pol}}    
\end{equation}

This method now correctly expresses the relative shear velocity of the background solar wind flowing tangentially to the flux rope magnetic field. The velocity vectors of this newly-defined tangential flow is displayed in figure \ref{2by2}a. The intensity of the $|\mathbf{v}_{rel_{\parallel}}|$ is displayed in figure \ref{2by2}d along with 2 contour lines: $|\mathbf{v}_{A_{pol}}| =20 km/s$ (white curve as shown in figure \ref{2by2}c), and $We_{\parallel}=1$ - i.e. using the velocity from equation \ref{eqVe} (grey curve). The newer Weber equipartition line ($We_{\parallel}$) surrounds a larger part of the FR than the original Weber line. $We_{\parallel}$ line displays an extremely extended region outside, but connected to, the FR leading edge and is out of the equatorial plane. This thin region is within the sheath region and approximately follows the curvature of the shock front. The reason this region is included within the equipartition line is similar to that of the extended regions of the We=1 boundary - namely the $|\mathbf{v}_{rel_{\parallel}}|$ is smaller than its surroundings.

\begin{figure}
\includegraphics[scale=0.55]{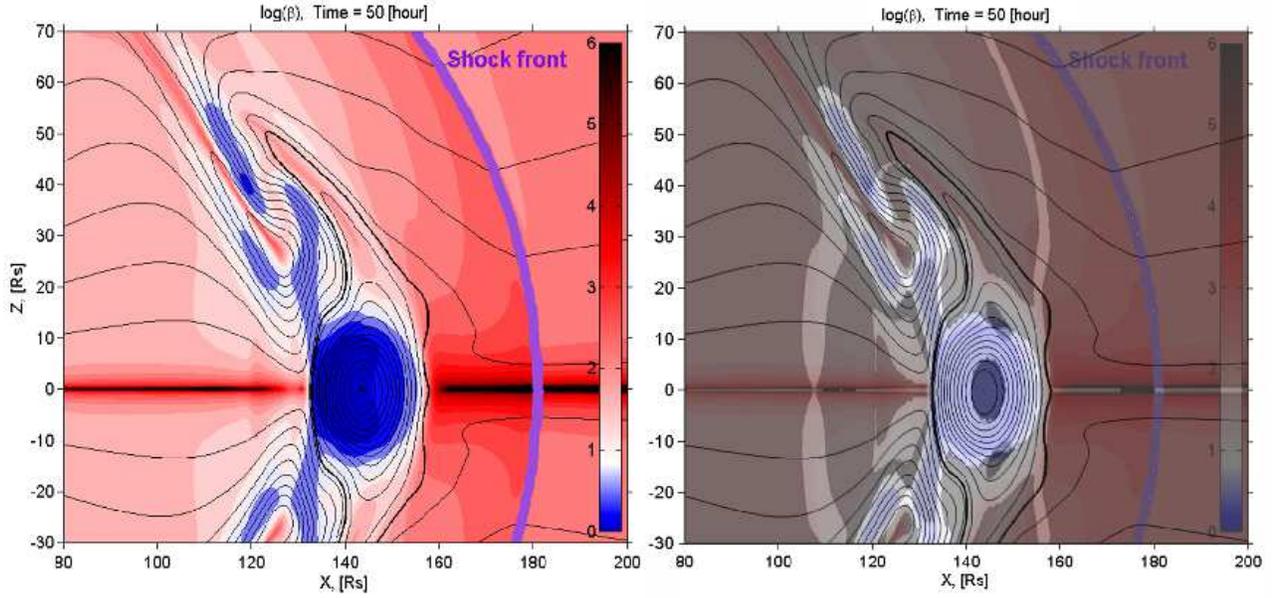}
\caption{(Left) Displays the plasma $\beta$ (colour scale) with position of the shock front for the same heliocentric distance as shown in figure \ref{2by2}. The magnetic flux function are shown as black contours with the thicker black line demarcating the outer boundary of the FR. (Right) The same as the left figure but with two additional semi-transparent layers overplotted. One of the semi-transparent layers displays a white region for $We_{\parallel}\leq 1$ and black otherwise. The other layer shows a white region for $|\mathbf{v}_{A_{pol}}| \geq 20 km/s$. Therefore the figure highlights the regions defined by the mathematical equipartition lines and darkens the surroundings.}
\label{weberNew}
\end{figure}

To understand how the sharp transition in plasma $\beta$ is related to the  different equipartition lines we display the intensity of the plasma $\beta$ in figure \ref{weberNew} using the same time frame as figure \ref{2by2}. The morphology of FR core in the figure, with a low plasma $\beta$, can be seen to be quasi-circular. Figure \ref{weberNew}b displays the same plasma $\beta$ intensity, but with 2 layers of semi-transparent grey shaded regions over-plotted. The boundary demarcating the light and dark shading of each layer are the same as those displayed in figure \ref{2by2}d. One of the semi-transparent layers displays a white region for $We_{\parallel}\leq 1$ and black otherwise. The other layer shows a white region for $|\mathbf{v}_{A_{pol}}| \geq 20 km/s$. Therefore figure \ref{weberNew}b highlights the regions defined by the mathematical equipartition lines and darkens the surroundings. We see that the $|\mathbf{v}_{A_{pol}}| = 20 km/s$ equipartition line, for the most part, tightly surrounds the low plasma $\beta$ core. The only exception is that of the thin extending section outside and to the rear of the FR. This region is the lightest region in the figure as it is also within the boundary of the second transparent layer. The region defined as $We_{\parallel}\leq 1$ but with $|\mathbf{v}_{A_{pol}}| \geq 20 km/s$ is the light grey region. Concentrating on this light grey region that remains within the magnetic flux rope, we notice that the area extends a little further in the `wing' sections of the FR. The boundary of the $We_{\parallel}= 1$ demarcates the region where the plasma $\beta$ has just transitioned into the higher values. Therefore the region inside the FR and between the two equipartition lines (light grey region) identifies the section where the sharp plasma beta transition occurs. Inside the transition region (i.e. closer to the FR center) the topology of the FR displays a quasi-cylindrical structure.

\section{Summary and Discussion}
Recent developments in heliospheric imagery from the STEREO mission have often shown that CMEs continue to propagate radially away from the Sun and deform into oblate shapes. This is further supported by heliospheric simulations such as the one often used in space weather forecasting, Enlil. However strong conflicting evidence from in situ data along one dimension inside CMEs suggest that they often have a magnetically circular cross-section. In this paper we hypothesise that a FR may form a magnetic substructure core as an explanation as to why remote observations sometimes may appear to contradict the topology suggested by in situ data. The Weber equipartition line defines the portion of the obstacle that will stay together. But there is a boundary layer surrounding the line where the plasma properties between the two regions smoothly but drastically change. This boundary layer is where the SW is able to drag the outer flux rope field lines away. Therefore the drastic plasma $\beta$ change is a consequence of the tension force not restraining the field at the outer edges.

\subsection{Interpretations for remote sensing}
Although our simulation would benefit from more advanced and realistic background solar wind conditions, we investigate the plasma $\beta$ within the CME and how this affects the distribution of flux. As the CME propagates the plasma $\beta$ transitions between low to high, which at first starts as a sharp boundary at the front edge of the FR structure (i.e. inner boundary of the sheath). This boundary then diffuses into a layer that surrounds the Weber equipartition line and moves within the FR field itself. This leads to a single CME behaving with two different properties: an inner core that is dominated by a stronger twist magnetic field which will favour a circular shape due to magnetic tension and the peripheral edge which has a stronger interaction with the ambient solar wind and is therefore prone to deformation. 

In a sense, the Weber equipartition line acts as a belt or `rubber band' around the core region, restraining flux from stretching out. But with the presence of a boundary layer surrounding the equipartition line, magnetic flux (field lines) are able to slowly leak through the Weber line to form part of the extend wings and then eventually disconnect from the flux rope all together. In our simulation we show that the flux rope core contains approximately $80\%$ of the original toroidal flux while remaining about $40\%$ of the overall size.

EM98 performed a similar simulation of a twisted flux tube within a high-plasma $\beta$ regime, akin to a tube rising in the solar convection zone, to show that the conversion of the tube into two vortex rolls can be suppressed by increasing the magnetic field twist. This work showed that the FR maintained its structure when conditions regarding the inflow velocity and poloidal Alfv\'en velocity were met. EM98 performed a parametric study by varying the amount of magnetic field twist within the flux rope. Such a parametric study would be a natural extension to the work presented here. Therefore further detailed studies which involve varying the tension force or changing the plasma interaction (by changing the background plasma density or plasma $\beta$) should be performed to better understand the extent to which the substructure occurs. A useful focus would be to understand whether this plasma $\beta$ transition scenario is a common occurrence or if it exists for only extreme (geoeffective) events. Another useful extension would be to perform a parametric study of CME speeds to better understand the cause of the complexity in the Weber Boundary for conditions with/without a propagating shock front. Understanding these thresholds may also provide clues into the requirements and limitations of magnetic field twist during the initial configuration above the solar surface. Developing more advanced simulations that realistically replicate the background solar wind in three dimensions would aid further understanding of this phenomena.

By injecting a CME instantaneously with a large magnetic flux, we emulate an extreme acceleration profile for a fast CME entering the inner heliosphere which is akin to that expected for a Carrington event or other extreme cases. In this simulation, the propagation speed of the flux rope through the heliosphere is more comparable to a typical fast CME. Under these conditions it is clear that a head-on impact onto the Earth will have a drastically different effect to one that, at first sight, would appear to be only moderately away from the Sun-Earth line. However further work could benefit from investigating a more physical acceleration process similar to \citet{shiota2005}. The initiation process of launching a FR by \citet{shiota2005} shows that a propagating CME might display the characteristic of two regimes separated by surface of $\beta=1$ (see Figure 13 within the authors work).

The Enlil model is often used for space weather forecasting. However this model is usually implemented by assuming the CME can be initially modelled as a high density blob with no internal magnetic field. The work in this paper suggests that without a restraining force from a twisted field the CME would indeed deform and spread out into a pancake structure. But interestingly, the magnetic field from the background solar wind often drapes around the CME during its propagation. This sheath region field could potentially serve as a restorative force preventing the CME from largely disintegrating by the time it arrives at Earth. Further work into understanding the extent to which draping field ahead of a CME can act as a restraining force should also be pursued.

At terrestrial distances, remote observations of CMEs using the HI-2 camera on board the STEREO mission have often identified flattened `pancaked' structures. However, this is commonly performed by locating a dark cavity or a bright `sheath' region that surrounds the probable flux rope \citep[e.g.][]{howardt2012b,deforest2013}. While the sheath region has an elongated structure due to the overall morphology of a magnetic flux rope, the imagery does not express the topological distribution of flux within the observed cavity.


\subsection{Interpretations for in situ data}
Recent articles from in situ analysis have alluded to the possibility that different parts of a single CME may have differing in situ properties. By using in situ data and a cylindrical model, \citet{dasso2005} investigated the flux content of 20 CMEs and showed the inner core ($\sim30\%$ of its total size) remained more symmetric. Later \citet{demoulin2009} created a flux rope model that relaxed the cylindrical symmetry explicitly to better estimate the cross sectional shape from in situ data. D{\'e}moulin et al. investigated elliptical shapes with a variety of eccentricities. The authors observed that the sharper (more deformed) parts of the boundary impose a stronger curvature, therefore the stronger magnetic tension reduces the field line bending inside the flux rope. For larger bending of the outer boundary, the magnetic tension increases such that the core field lines slightly shrink towards the center, forming a more circular shape. 

In addition, for elliptically shaped flux ropes with significant deformations, the authors showed that the magnetic pressure is lower at the flux rope sides than the center. Assuming the plasma pressure remains uniform, they argue that even for very low plasma $\beta$, the force-free approximation is no longer valid. They therefore expect these limb (`wing') regions to be swept away by the solar wind. They further hypothesis that reconnection with the solar wind magnetic field will contribute to removing these parts, while the core field maintains its identity.

\citet{dasso2012} also highlighted this conflict by hypothesising the magnetic tension may favour a circular shape and interpreting the external part as an interaction with the ambient solar wind. Our simulation support such a scenario of a core flux rope CME which is significantly smaller than the full extent of the original flux rope. Therefore if a spacecraft were to travel through the center of a flux rope, the in situ magnetic field measurements made would suggest a quasi-cylindrical obstacle similar to the core substructure shown in our simulations. 

Several authors have previously shown schematics suggesting that a quasi-cylindrical FR may be part of a larger magnetic obstacle \citep[e.g.][]{jian2006, riley2013}. Thus the trajectory of a spacecraft will dictate whether a cylindrical FR topology is measured or not. This then brings into question whether all CMEs are in fact FRs or not \citep{vourlidas2012} - or more accurately speaking, if a FR topology always exists somewhere within the entire CME structure. \citet{riley2013} investigated the plasma $\beta$ properties of 181 CMEs using a list of events provided by \citet{du2010}. \citet{riley2013} showed that the coherent structure of in situ measured FRs often have a significantly lower plasma $\beta$ value than events without a FR topology (see Figure 3 of their paper). These authors also find that the plasma $\beta$ is one of the most significant predictor variables for the presence of a FR topology. The Weber equipartition analysis in this paper provides a scientific explanation to their result.

Several other interpretations for creating substructures within a CME which ultimately leads to a quasi-cylindrical flux rope at terrestrial distances have been proposed. For example, \cite{owens2009} proposed large current sheets can lead to magnetic reconnection which, in an extreme case, could lead towards forming multiple small flux ropes from a single larger one. A review of other potential scenarios for creating substructures can be found in \citet{steed2011}.

If a CME contains a magnetic substructure which is significantly more cylindrical than the entire morphology, then in situ measurements should show CME-like properties outside the magnetic flux rope as indicated by a multi-spacecraft analysis by \citet{reinard2012}. A study by \cite{richardson2010} investigated CMEs over solar cycle 23 and defined boundaries by the ionisation charge states of certain heavy ions. Figure 2 and 3 from their work shows that this may be a frequent occurrence. However the charge states do not always overlap with the magnetic signatures \cite[e.g.][]{rakowski2011,lepri2012} and so can not be used as a definitive explanation.  Therefore it is important for future work to investigate simulated time-series at different locations within the CME \cite[similar to][]{riley2004b}.


\acknowledgments
N.P.S. was partially supported by the NASA Living With a Star Jack Eddy Postdoctoral Fellowship Program, administered by the UCAR Visiting Scientist Programs and hosted by the Naval Research Laboratory. A.V. and M.G.L. were supported by NASA and the office of Naval Research. The numerical calculations were performed using the supercomputing cluster at the Solar-Terrestrial Environment Laboratory, Nagoya University.

\bibliographystyle{apj}

\begin{thebibliography}{57}
\expandafter\ifx\csname natexlab\endcsname\relax\def\natexlab#1{#1}\fi

\bibitem[{{Barnard} {et~al.}(2011){Barnard}, {Lockwood}, {Hapgood}, {Owens},
  {Davis}, \& {Steinhilber}}]{barnard2011}
{Barnard}, L., {Lockwood}, M., {Hapgood}, M.~A., {et~al.} 2011, \grl, 38, 16103

\bibitem[{{Blanco} {et~al.}(2013){Blanco}, {Hidalgo}, {G{\'o}mez-Herrero},
  {Rodr{\'{\i}}guez-Pacheco}, {Heber}, {Wimmer-Schweingruber}, \&
  {Mart{\'{\i}}n}}]{blanco2013}
{Blanco}, J.~J., {Hidalgo}, M.~A., {G{\'o}mez-Herrero}, R., {et~al.} 2013,
  \aap, 556, A146

\bibitem[{{Burlaga} {et~al.}(1981){Burlaga}, {Sittler}, {Mariani}, \&
  {Schwenn}}]{burlaga1981}
{Burlaga}, L., {Sittler}, E., {Mariani}, F., \& {Schwenn}, R. 1981, \jgr, 86,
  6673

\bibitem[{{Dasso} {et~al.}(2012){Dasso}, {D{\'e}moulin}, \&
  {Gulisano}}]{dasso2012}
{Dasso}, S., {D{\'e}moulin}, P., \& {Gulisano}, A.~M. 2012, in IAU Symposium,
  Vol. 286, IAU Symposium, ed. C.~H. {Mandrini} \& D.~F. {Webb}, 139--148

\bibitem[{{Dasso} {et~al.}(2005){Dasso}, {Gulisano}, {Mandrini}, \&
  {D{\'e}moulin}}]{dasso2005}
{Dasso}, S., {Gulisano}, A.~M., {Mandrini}, C.~H., \& {D{\'e}moulin}, P. 2005,
  Advances in Space Research, 35, 2172

\bibitem[{{DeForest} {et~al.}(2013){DeForest}, {Howard}, \&
  {McComas}}]{deforest2013}
{DeForest}, C.~E., {Howard}, T.~A., \& {McComas}, D.~J. 2013, \apj, 769, 43

\bibitem[{{D{\'e}moulin} \& {Dasso}(2009)}]{demoulin2009}
{D{\'e}moulin}, P., \& {Dasso}, S. 2009, \aap, 507, 969

\bibitem[{{D{\'e}moulin} {et~al.}(2002){D{\'e}moulin}, {Mandrini}, {van
  Driel-Gesztelyi}, {Thompson}, {Plunkett}, {Kov{\'a}ri}, {Aulanier}, \&
  {Young}}]{demoulin2002}
{D{\'e}moulin}, P., {Mandrini}, C.~H., {van Driel-Gesztelyi}, L., {et~al.}
  2002, \aap, 382, 650

\bibitem[{{Du} {et~al.}(2010){Du}, {Zuo}, \& {Zhang}}]{du2010}
{Du}, D., {Zuo}, P.~B., \& {Zhang}, X.~X. 2010, \solphys, 262, 171

\bibitem[{{Emonet} \& {Moreno-Insertis}(1998)}]{emonet1998}
{Emonet}, T., \& {Moreno-Insertis}, F. 1998, \apj, 492, 804

\bibitem[{{Hidalgo} {et~al.}(2002){Hidalgo}, {Nieves-Chinchilla}, \&
  {Cid}}]{hidalgo2002}
{Hidalgo}, M.~A., {Nieves-Chinchilla}, T., \& {Cid}, C. 2002, \grl, 29, 130000

\bibitem[{{Howard} {et~al.}(2008){Howard}, {Moses}, {Vourlidas}, {Newmark},
  {Socker}, {Plunkett}, {Korendyke}, {Cook}, {Hurley}, {Davila}, {Thompson},
  {St Cyr}, {Mentzell}, {Mehalick}, {Lemen}, {Wuelser}, {Duncan}, {Tarbell},
  {Wolfson}, {Moore}, {Harrison}, {Waltham}, {Lang}, {Davis}, {Eyles},
  {Mapson-Menard}, {Simnett}, {Halain}, {Defise}, {Mazy}, {Rochus}, {Mercier},
  {Ravet}, {Delmotte}, {Auchere}, {Delaboudiniere}, {Bothmer}, {Deutsch},
  {Wang}, {Rich}, {Cooper}, {Stephens}, {Maahs}, {Baugh}, {McMullin}, \&
  {Carter}}]{howard2008}
{Howard}, R.~A., {Moses}, J.~D., {Vourlidas}, A., {et~al.} 2008, Space Science
  Review, 136, 67

\bibitem[{{Howard} \& {DeForest}(2012)}]{howardt2012b}
{Howard}, T.~A., \& {DeForest}, C.~E. 2012, \apj, 746, 64

\bibitem[{{Hu} \& {Sonnerup}(2002)}]{hu2002}
{Hu}, Q., \& {Sonnerup}, B.~U.~{\"O}. 2002, Journal of Geophysical Research
  (Space Physics), 107, 1142

\bibitem[{{Jian} {et~al.}(2006){Jian}, {Russell}, {Luhmann}, \&
  {Skoug}}]{jian2006}
{Jian}, L., {Russell}, C.~T., {Luhmann}, J.~G., \& {Skoug}, R.~M. 2006,
  \solphys, 239, 393

\bibitem[{{Lepping} {et~al.}(1990){Lepping}, {Burlaga}, \&
  {Jones}}]{lepping1990}
{Lepping}, R.~P., {Burlaga}, L.~F., \& {Jones}, J.~A. 1990, \jgr, 95, 11957

\bibitem[{{Lepri} {et~al.}(2012){Lepri}, {Laming}, {Rakowski}, \& {von
  Steiger}}]{lepri2012}
{Lepri}, S.~T., {Laming}, J.~M., {Rakowski}, C.~E., \& {von Steiger}, R. 2012,
  \apj, 760, 105

\bibitem[{{Liu} {et~al.}(2008){Liu}, {Luhmann}, {Huttunen}, {Lin}, {Bale},
  {Russell}, \& {Galvin}}]{liu2008}
{Liu}, Y., {Luhmann}, J.~G., {Huttunen}, K.~E.~J., {et~al.} 2008, \apjl, 677,
  L133

\bibitem[{{Liu} {et~al.}(2006){Liu}, {Richardson}, {Belcher}, {Wang}, {Hu}, \&
  {Kasper}}]{liu2006}
{Liu}, Y., {Richardson}, J.~D., {Belcher}, J.~W., {et~al.} 2006, Journal of
  Geophysical Research (Space Physics), 111, 12

\bibitem[{{Lockwood} {et~al.}(2004){Lockwood}, {Forsyth}, {Balogh}, \&
  {McComas}}]{lockwood2004}
{Lockwood}, M., {Forsyth}, R., {Balogh}, A., \& {McComas}, D. 2004, Annales
  Geophysicae, 22, 1395

\bibitem[{{Lockwood} \& {Owens}(2011)}]{lockwood2011}
{Lockwood}, M., \& {Owens}, M.~J. 2011, Journal of Geophysical Research (Space
  Physics), 116, 4109

\bibitem[{{Lugaz} {et~al.}(2005){Lugaz}, {Manchester}, \&
  {Gombosi}}]{lugaz2005}
{Lugaz}, N., {Manchester}, IV, W.~B., \& {Gombosi}, T.~I. 2005, \apj, 627, 1019

\bibitem[{{Lundquist}(1950)}]{lundquist1950}
{Lundquist}, S. 1950, Ark. Fys., 2

\bibitem[{{Lynch} {et~al.}(2005){Lynch}, {Gruesbeck}, {Zurbuchen}, \&
  {Antiochos}}]{lynch2005}
{Lynch}, B.~J., {Gruesbeck}, J.~R., {Zurbuchen}, T.~H., \& {Antiochos}, S.~K.
  2005, Journal of Geophysical Research (Space Physics), 110, 8107

\bibitem[{Marubashi(1997)}]{marubashi1997}
Marubashi, K. 1997, in Geophys. Monogr. Ser., Vol.~99, Coronal Mass Ejections,
  ed. N.~{Crooker}, J.~A. {Joselyn}, \& J.~{Feynman} (AGU, Washington, D. C.),
  147--156

\bibitem[{{Marubashi}(2000)}]{marubashi2000}
{Marubashi}, K. 2000, Advances in Space Research, 26, 55

\bibitem[{{M{\"o}stl} {et~al.}(2009){M{\"o}stl}, {Farrugia}, {Biernat},
  {Leitner}, {Kilpua}, {Galvin}, \& {Luhmann}}]{mostl2009a}
{M{\"o}stl}, C., {Farrugia}, C.~J., {Biernat}, H.~K., {et~al.} 2009, \solphys,
  256, 427

\bibitem[{{Nieves-Chinchilla} {et~al.}(2012){Nieves-Chinchilla}, {Colaninno},
  {Vourlidas}, {Szabo}, {Lepping}, {Boardsen}, {Anderson}, \&
  {Korth}}]{nieves2012}
{Nieves-Chinchilla}, T., {Colaninno}, R., {Vourlidas}, A., {et~al.} 2012,
  Journal of Geophysical Research (Space Physics), 117, 6106

\bibitem[{{Odstrcil} \& {Pizzo}(1999)}]{odstrcil1999}
{Odstrcil}, D., \& {Pizzo}, V.~J. 1999, \jgr, 104, 28225

\bibitem[{{Owens}(2006)}]{owens2006b}
{Owens}, M.~J. 2006, Journal of Geophysical Research (Space Physics), 111,
  12109

\bibitem[{{Owens}(2009)}]{owens2009}
---. 2009, \solphys, 260, 207

\bibitem[{{Owens} \& {Crooker}(2006)}]{owens2006c}
{Owens}, M.~J., \& {Crooker}, N.~U. 2006, Journal of Geophysical Research
  (Space Physics), 111, 10104

\bibitem[{{Owens} {et~al.}(2008){Owens}, {Crooker}, {Schwadron}, {Horbury},
  {Yashiro}, {Xie}, {St.~Cyr}, \& {Gopalswamy}}]{owens2008}
{Owens}, M.~J., {Crooker}, N.~U., {Schwadron}, N.~A., {et~al.} 2008, \grl, 35,
  20108

\bibitem[{{Owens} {et~al.}(2012){Owens}, {D{\'e}moulin}, {Savani}, {Lavraud},
  \& {Ruffenach}}]{owens2012}
{Owens}, M.~J., {D{\'e}moulin}, P., {Savani}, N.~P., {Lavraud}, B., \&
  {Ruffenach}, A. 2012, Solar Physics, 278, 435

\bibitem[{{Owens} {et~al.}(2006){Owens}, {Merkin}, \& {Riley}}]{owens2006}
{Owens}, M.~J., {Merkin}, V.~G., \& {Riley}, P. 2006, Journal of Geophysical
  Research (Space Physics), 111, 3104

\bibitem[{{Rakowski} {et~al.}(2011){Rakowski}, {Laming}, \&
  {Lyutikov}}]{rakowski2011}
{Rakowski}, C.~E., {Laming}, J.~M., \& {Lyutikov}, M. 2011, \apj, 730, 30

\bibitem[{{Reinard} {et~al.}(2012){Reinard}, {Lynch}, \&
  {Mulligan}}]{reinard2012}
{Reinard}, A.~A., {Lynch}, B.~J., \& {Mulligan}, T. 2012, \apj, 761, 175

\bibitem[{{Richardson} \& {Cane}(2010)}]{richardson2010}
{Richardson}, I.~G., \& {Cane}, H.~V. 2010, \solphys, 264, 189

\bibitem[{{Riley} \& {Crooker}(2004)}]{riley2004}
{Riley}, P., \& {Crooker}, N.~U. 2004, \apj, 600, 1035

\bibitem[{{Riley} \& {Richardson}(2013)}]{riley2013}
{Riley}, P., \& {Richardson}, I.~G. 2013, \solphys, 284, 217

\bibitem[{{Riley} {et~al.}(2004){Riley}, {Linker}, {Lionello}, {Miki{\'c}},
  {Odstrcil}, {Hidalgo}, {Cid}, {Hu}, {Lepping}, {Lynch}, \&
  {Rees}}]{riley2004b}
{Riley}, P., {Linker}, J.~A., {Lionello}, R., {et~al.} 2004, Journal of
  Atmospheric and Solar-Terrestrial Physics, 66, 1321

\bibitem[{{Savani} {et~al.}(2010){Savani}, {Owens}, {Rouillard}, {Forsyth}, \&
  {Davies}}]{savani2010}
{Savani}, N.~P., {Owens}, M.~J., {Rouillard}, A.~P., {Forsyth}, R.~J., \&
  {Davies}, J.~A. 2010, \apjl, 714, L128

\bibitem[{{Savani} {et~al.}(2011{\natexlab{a}}){Savani}, {Owens}, {Rouillard},
  {Forsyth}, {Kusano}, {Shiota}, \& {Kataoka}}]{savani2011a}
{Savani}, N.~P., {Owens}, M.~J., {Rouillard}, A.~P., {et~al.}
  2011{\natexlab{a}}, \apj, 731, 109

\bibitem[{{Savani} {et~al.}(2009){Savani}, {Rouillard}, {Davies}, {Owens},
  {Forsyth}, {Davis}, \& {Harrison}}]{savani2009}
{Savani}, N.~P., {Rouillard}, A.~P., {Davies}, J.~A., {et~al.} 2009, Annales
  Geophysicae, 27, 4349

\bibitem[{{Savani} {et~al.}(2012){Savani}, {Shiota}, {Kusano}, {Vourlidas}, \&
  {Lugaz}}]{savani2012b}
{Savani}, N.~P., {Shiota}, D., {Kusano}, K., {Vourlidas}, A., \& {Lugaz}, N.
  2012, \apj, 759, 103

\bibitem[{{Savani} {et~al.}(2013){Savani}, {Vourlidas}, {Pulkkinen},
  {Nieves-Chinchilla}, {Lavraud}, \& {Owens}}]{savani2013}
{Savani}, N.~P., {Vourlidas}, A., {Pulkkinen}, A., {et~al.} 2013, Space
  Weather, 11, 245

\bibitem[{{Savani} {et~al.}(2011{\natexlab{b}}){Savani}, {Owens}, {Rouillard},
  {Forsyth}, {Kusano}, {Shiota}, {Kataoka}, {Jian}, \& {Bothmer}}]{savani2011b}
{Savani}, N.~P., {Owens}, M.~J., {Rouillard}, A.~P., {et~al.}
  2011{\natexlab{b}}, \apj, 732, 117

\bibitem[{{Schreiner} {et~al.}(2012){Schreiner}, {Cattell}, {Kersten}, \&
  {Hupach}}]{schreiner2012}
{Schreiner}, S., {Cattell}, C., {Kersten}, K., \& {Hupach}, A. 2012, \solphys,
  19

\bibitem[{{Shiota} {et~al.}(2005){Shiota}, {Isobe}, {Chen}, {Yamamoto},
  {Sakajiri}, \& {Shibata}}]{shiota2005}
{Shiota}, D., {Isobe}, H., {Chen}, P.~F., {et~al.} 2005, \apj, 634, 663

\bibitem[{{Shiota} {et~al.}(2008){Shiota}, {Kusano}, {Miyoshi}, {Nishikawa}, \&
  {Shibata}}]{shiota2008}
{Shiota}, D., {Kusano}, K., {Miyoshi}, T., {Nishikawa}, N., \& {Shibata}, K.
  2008, Journal of Geophysical Research (Space Physics), 113, 3

\bibitem[{{Steed} {et~al.}(2011){Steed}, {Owen}, {D{\'e}moulin}, \&
  {Dasso}}]{steed2011}
{Steed}, K., {Owen}, C.~J., {D{\'e}moulin}, P., \& {Dasso}, S. 2011, Journal of
  Geophysical Research (Space Physics), 116, 1106

\bibitem[{{Suess}(1988)}]{suess1988}
{Suess}, S.~T. 1988, \jgr, 93, 5437

\bibitem[{{Thernisien}(2011)}]{thernisien2011}
{Thernisien}, A. 2011, \apjs, 194, 33

\bibitem[{{Thernisien} {et~al.}(2009){Thernisien}, {Vourlidas}, \&
  {Howard}}]{thernisien2009}
{Thernisien}, A., {Vourlidas}, A., \& {Howard}, R.~A. 2009, Solar Physics, 256,
  111

\bibitem[{{Tsurutani} \& {Gonzalez}(1997)}]{tsurutani1997}
{Tsurutani}, B.~T., \& {Gonzalez}, W.~D. 1997, in Washington DC American
  Geophysical Union Geophysical Monograph Series, Vol.~98, Washington DC
  American Geophysical Union Geophysical Monograph Series, ed. B.~T.
  {Tsurutani}, W.~D. {Gonzalez}, Y.~{Kamide}, \& J.~K. {Arballo}, 77--89

\bibitem[{{Vourlidas} {et~al.}(2012){Vourlidas}, {Lynch}, {Howard}, \&
  {Li}}]{vourlidas2012}
{Vourlidas}, A., {Lynch}, B.~J., {Howard}, R.~A., \& {Li}, Y. 2012, \solphys,
  192

\bibitem[{{Wood} {et~al.}(2012){Wood}, {Rouillard}, {M{\"o}stl}, {Battams},
  {Savani}, {Marubashi}, {Howard}, \& {Socker}}]{wood2012b}
{Wood}, B.~E., {Rouillard}, A.~P., {M{\"o}stl}, C., {et~al.} 2012, Solar
  Physics, 143

\end{thebibliography}

\clearpage


\clearpage

\clearpage

\clearpage

\clearpage

\clearpage

\end{document}